\documentclass[10pt,twoside]{article}
\usepackage{epsfig,amsmath,amssymb,verbatim}

\begin{document}
\setcounter{page}{1}

\title{Characterization and Mitigation of Sinusoidal Carrier Modulation Signals by Short-Term Sinusoidal Analysis} 

\author{S.W.\ Ellingson\thanks{Bradley Dept.\ of Electrical \& Computer Engineering, 302 Whittemore Hall, Virginia Polytechnic Institute \& State University, Blacksburg VA 24061 USA. E-mail: {\tt ellingson.1@vt.edu}.} 
}

\maketitle



\begin{abstract}
Most radio signals can be described as a form of sinusoidal carrier modulation.
These signals have the property that when observed over a sufficiently short period of time, the waveform is approximately sinusoidal, exhibiting constant magnitude, frequency, and phase.  
When coherently digitized with sufficiently high sample rate, it is possible to discern this sinusoid. 
The associated signal can be represented as a sinusoid with time-varying frequency, magnitude, and phase.
These parameters can subsequently be used for detection and characterization, including identification.
The parameters can also be used to synthesize a noise-free estimate of the signal.
The synthesized signal can be coherently subtracted from the original data in order to cancel the signal, nominally without distortion of the associated noise or nearby signals.
In this report, the conditions required for the short-time sinusoidal approximation are described.
Also, performance of the canceling technique is examined in the case of a VHF-band narrowband FM-modulated signal. 
\end{abstract}


\section{Introduction}
\label{sIntro}

Many radio signals are well-described as sinusoidal carrier modulation (SCM).
Examples include most wireless communications signals and many unintentional signals of anthropogenic origin. 
SCM has the property that when observed over a sufficiently short period of time, the associated waveform is approximately sinusoidal; i.e., is approximately a sinusoid having constant magnitude, frequency, and phase.  
When such signals are coherently digitized with sufficiently high sample rate, it is possible to discern the associated short-term sinusoidal waveform.  
In this report, we refer to this as ``short-term sinusoidal analysis'' (STSA).
The short-term sinusoidal condition is frequently met in high-performance wideband receivers used in a variety of applications including electronic warfare and radio astronomy.

Estimates of magnitude, frequency, and phase of the short-term sinusoid over time can be used to synthesize an estimate of the signal waveform over longer periods.
This estimate is nominally noise-free and may subsequently be used for detection and characterization, including identification.
The estimated waveform may also be coherently subtracted from the original data in order to cancel the signal.

This approach is not new: A similar strategy is described in \cite{MPM98}. 
In the present report, the focus is on quantifying the conditions required for the short-time sinusoidal approximation to be valid, and examining performance considering the full bandwidth of the signal. 
This is an important consideration in certain applications, such as radio astronomy, where even the relatively weak edges of the interfering signal's spectrum can overwhelm underlying signals of interest.
The principal advantage of the STSA-based canceler relative to other methods for coherent subtraction of interference in radio astronomy (e.g., \cite{EBB01}, \cite{EH03}, and references therein), is that the STSA-based canceler requires no advance knowledge of the modulation. 

In this report, an STSA-based canceler is applied to data containing narrowband FM radio signals captured off-the-air.
The signals originate from NOAA Weather Radio (NWR) stations at about 162~MHz \cite{NWR}, having bandwidth $\approx$10~kHz, captured at 2.048~MSPS using 8-bit ``I'' $+$ 8-bit ``Q'' complex sampling.  
Precise estimation is demonstrated, including the ability to coherently suppress this signal by about 17~dB by subtracting the estimated waveform from the original data.
This suppression is achieved with negligible distortion to the remaining signals and noise, and evidence indicates that up to 18~dB of additional suppression is possible with enhancements described in this report. 

The remainder of this report is organized as follows:
Section~\ref{sTheory} explains the relevant theory.
Section~\ref{sAlgorithm} (``Algorithm Considered in this Study'') describes the specific implementation of STSA employed in this study.
Section~\ref{sResults} (``Results'') reports a demonstration of the application of this implementation of STSA to the NWR signal.
This report concludes with a discussion of applications of this method in Section~\ref{sApplications} and ideas for future work in Section~\ref{sFW}.

\section{Theory}
\label{sTheory}

A signal $x(t)$ exhibiting SCM can be described in complex baseband representation as follows:
\begin{equation}
x(t) = A(t) ~e^{j\left[\omega(t)t + \psi(t)\right]}
\label{eSCM}
\end{equation}
where $j=\sqrt{-1}$ and $A(t)$, $\omega(t)$, and $\psi(t)$ are the magnitude, frequency, and phase of the signal as a function of time.
SCMs include 
amplitude modulation (AM), which consists of varying only $A$;
frequency modulation (FM), which consists of varying only $\omega$; and
phase modulation (PM), which consists of varying only $\psi$.
The corresponding digital modulations
amplitude shift keying (ASK),
frequency shift keying (FSK), and
phase shift keying (PSK) are representable using Equation~\ref{eSCM} without modification, even when pulse shaping is employed.
Quadrature-amplitude modulation (QAM), which is implemented by simultaneously varying both $A$ and $\psi$, is similarly representable using Equation~\ref{eSCM}.
The list of applicable modulations continues; for additional background on this topic, Chapters~5 and 6 of \cite{E16} are suggested. 

\subsection{Short-Term Sinusoidal Condition}
\label{sSTSC}

The spectrum of a SCM signal depends on which of the parameters $A$, $\omega$, and $\psi$ are varied and the manner and rate in which they are varied.
Therefore different modulations exhibit different spectra. 
However if one observes $x(t)$ over a sufficiently short time, no variation will be apparent.
That is, the signal will appear, approximately, to be a sinusoid having \emph{constant} parameters $A$, $\omega$, $\psi$.
This condition exists when
\begin{equation}
T \lessapprox \frac{1}{B}
\end{equation}
where $T$ is the observation interval, and $B$ is the bandwidth of $x(t)$ as determined from the time-average power spectral density.
To exploit this condition, the sample rate $F_S$ of the data containing $x(t)$ must sufficiently large 
to be able to resolve $x(t)$ from other signals that may be simultaneously present in the passband.  
Let $N$ be the number of samples taken over time $T$.  
Then $T=N/F_S$ and subsequently 
\begin{equation}
F_S \gtrapprox NB
\end{equation}

Also, note that it is assumed that data sampled at this rate have bandwidth at or near the Nyquist bandwidth $F_S/2$.  
For example, if coherently-sampled data are filtered to an effective bandwidth much less than $F_S/2$ before or after sampling, then the signal has been ``oversampled'' from a signal processing perspective and the effective sample rate for our purposes is much less than the actual sample rate.  That is, the same data could be perfectly represented at a lower sample rate, so the short-term sinusoidal condition applies to this lower sample rate. 

Let us consider the case to be demonstrated in Section~\ref{sResults} as an example.
The NWR narrowband FM signals exhibit $B\approx 10$~kHz, the data are sampled at $F_S=2.048$~MSPS (complex), and data bandwidth is approximately Nyquist-limited; i.e., approximately 2~MHz. 
Therefore we require $T \lessapprox 100~\mu$s and subsequently $N \lessapprox 204$.
Generally the largest suitable values of $T$ and $N$ are desirable, since increasing $T$ improves estimation performance and increasing $N$ facilitates greater spectral resolution, which may be needed to resolve and identify individual signals.
In fact, choosing $N=F_S/B$ will usually allow the frequency ($\omega$) to be determined to an accuracy of $B/10$ to $B/100$ depending on other factors.\footnote{Note that a fast Fourier transform (FFT) would be limited to resolution of only $B$. Here we refer to the theoretical limit, which requires the use of a periodogram (as proposed in Section~\ref{ssOESS}) or a parametric technique.}  
This is important and required in order to be able to accurately track variation in $\omega$ with time. 

\subsection{Optimal Estimation of a Stationary Sinusoid}
\label{ssOESS}

In this report, we refer to a sinusoid whose parameters $A$, $\omega$, and $\psi$ remain constant over some interval $T$ as being ``stationary'' over that interval.
Optimal estimation of a single stationary sinusoid in additive white Gaussian noise is achieved as follows.
First, let us formally define the estimated waveform as follows:
\begin{equation}
\hat{x}(\hat{A},\hat{\omega},\hat{\psi};t) = \hat{A}e^{j\left(\hat{\omega}t+\hat{\psi}\right)}
\end{equation}
where $\hat{A}$, $\hat{\omega}$, and $\hat{\psi}$ are the corresponding estimated parameters.
Let us presume that the waveform of interest is embedded in noisy data $y(t)$ as follows:
\begin{equation}
y(t) = x(t) + n(t)
\end{equation}
where $n(t)$ is zero-mean Gaussian-distributed white noise.
In this case the frequency is optimally estimated as follows \cite{Kay93}.  
First, one evaluates 
\begin{equation}
\underset{\hat{\omega}}{\mathrm{arg\,max}} \left| \int_{t_0}^{t_0+T} y(t)~\hat{x}^*(1,\hat{\omega},0;t)~dt \right|
\label{ePeriodogram}
\end{equation}
This is simply correlation of the noisy data with a candidate waveform for all possible values of $\hat{\omega}$, and choosing the one which maximizes the correlation.  For this operation, $\hat{A}$ and $\hat{\psi}$ are irrelevant, so values of $1$ and $0$ are arbitrarily selected.
Once $\hat{\omega}$ is determined, the remaining parameters are determined as follows:
\begin{equation}
\hat{A}e^{j\hat{\psi}} = \int_{t_0}^{t_0+T} y(t)~\hat{x}^*(1,\hat{\omega},0;t)~dt
\label{eCorrelation}
\end{equation}
In other words, estimates of the magnitude and phase are given by the correlation with respect to $\hat{x}(1,\hat{\omega},0;t)$.  

\subsection{Estimation of a Stationary Sinusoid in the Presence of Other Sinusoids}
\label{ssESSPOS}

The procedure described in the previous section works well if $x(t)$ is the only signal present, and $T$ is sufficiently long to suppress noise.
If there are other signals present, and especially if they persist over the entire interval $T$, then the procedure will yield biased estimates.
In this case the optimal procedure is a modification of the single-tone procedure in which $M$ sinusoids are estimated jointly \cite{Kay93}.
This dramatically increases computational burden, and introduces the additional problem of how to determine the number of sinusoids $M$ (known as the ``model order problem'').
Determining $M$ is particularly difficult when the signal-to-noise ratio (SNR) is low, and error in determining $M$ leads to large biases in the subsequent estimation of waveform parameters.  
An alternative method which is sub-optimal but more robust is described below. 

The principal concept in this alternative method is that the bias is small as long as other signals are spectrally-resolved (i.e., separated by at least $F_S/N$ in frequency) and not too large relative to the signal of interest.  Bias can be further mitigated in this case by taking the following two measures.

First, applying a window to the length-$N$ block of samples before analysis will dramatically reduce bias.  This is because most of the bias is attributable to the incomplete periods at the beginning and the end of the length-$T$ sample block.  These ``endpoint'' contributions dominate the bias because correlation over the integer number of periods of a sinusoid interior to the block will be approximately zero, leaving the contributions from the incomplete periods at the beginning and end to dominate.  The optimal window depends on the degree to which $F_S$ is larger than $B$.  
For $F_S/B \approx N$, a triangular window is optimal because the triangular window yields minimum degradation of spectral resolution.      
For $F_S/B \gg N$, signals are spectrally well-resolved, in which case the excess spectral resolution can be traded off for the lower sidelobes offered by an alternative window such as the Hamming window.

The second measure is to identify and estimate the largest signal first.  This signal may then be subtracted from the data, which will reduce the contribution of this signal to the estimation bias for weaker signals.  In other words, estimate signals one at a time, from strongest to weakest, subtracting estimated waveforms as they are identified.

A variety of enhancements and hybrid methods can be conceived for dealing more effectively with the multiple-signal issue. However -- as demonstrated in Section~\ref{sResults} -- much can be accomplished simply by (1) windowing, and (2) analyzing signals in decreasing order of strength, as suggested above.

\subsection{Mitigation of Block-Boundary Discontinuities}
\label{sMitBlockBoundaryDisc}

The procedure described so far for estimation of stationary sinusoids presumes constant parameters $\hat{A}$, $\hat{\omega}$, and $\hat{\psi}$ over a block of data of length $T$.  These parameters are estimated independently for each subsequent block.  Therefore the presumed waveform will have a discontinuity at the block boundaries.  Aside from improperly modeling the actual variation of the parameters over time, the periodicity of the block-boundary discontinuity will create a spectral artifact appearing as a comb of tones with spacing $1/T$ in frequency.  Both problems may be addressed by interpolation of the waveforms.  The rudimentary scheme proposed and demonstrated in this report presumes the estimated waveform is nominal in the center of its block.  The waveform at all other times is obtained by linear interpolation of the present block's estimated waveform with that of the adjacent block.  This results in a continuous waveform, and one whose parameters are equal to those originally estimated for the block at the center of the block.  Crude it may be, this scheme turns out to be surprisingly effective.  Nevertheless, limitations are apparent in the results shown in Section~\ref{sResults} and possible enhancements are suggested in Section~\ref{sFW}.

\section{Algorithm Considered in this Study}
\label{sAlgorithm}

Based on the considerations presented in Section~\ref{sTheory}, here is a simple implementation of the STSA-based waveform estimator:
~\\

\noindent
For each length-$T$ block of $N$ samples:
\begin{enumerate}
\item \label{stWindow} Apply a triangular window.
\item FFT the data. If no peak exceeding a specified SNR threshold is identified, then exit.  Otherwise, identify the center frequency $f_c$ of the FFT bin containing the largest magnitude.
\item Use Equation~\ref{ePeriodogram} to determine $\hat{\omega}$.  
Search over uniformly-spaced frequencies separated by 1\% of the bin width, which is moderately finer than the finest resolution that can normally be expected from a periodogram-based procedure.
\item Use Equation~\ref{eCorrelation} to determine $\hat{A}$ and $\hat{\psi}$.
\item $\hat{A}\leftarrow\hat{A}/2$ to account for the triangular window applied in Step~\ref{stWindow}.  Save these values of the parameters for later use.
\item Subtract the estimated sinusoid from a copy of the original data.
\item Return to Step~\ref{stWindow}, now operating on the modified data from the previous step.
\end{enumerate}

Once parameters are available for two consecutive blocks, the estimated waveforms are constructed by linear interpolation as proposed in Section~\ref{sMitBlockBoundaryDisc}.  
These waveforms may subsequently be used for detection and/or characterization.
Alternatively these waveforms may be subtracted from the original data if mitigation is desired.  

\section{Results}
\label{sResults}

\begin{figure}
\begin{center}
\centering
\psfig{file=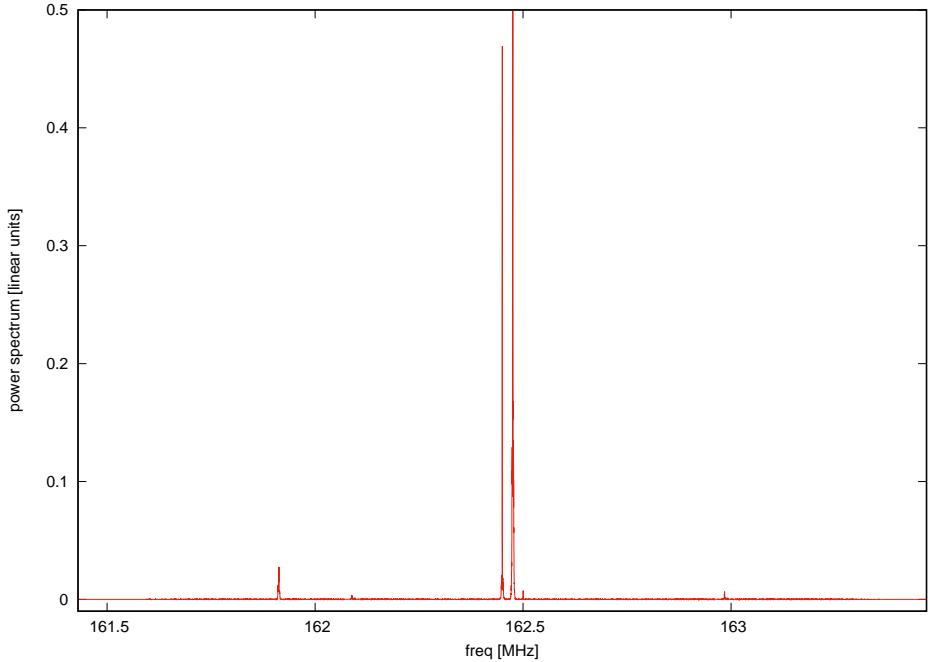,width=5in}
\end{center}
\caption{\label{fPowerSpectrum} Power spectrum of the original data in linear units. The peak of the 162.475~MHz signal is about 15 units and is truncated in order to more clearly show weaker signals.  Spectral resolution in this figure is 125~Hz.}
\end{figure}

In this section we demonstrate the use of the algorithm described in Section~\ref{sAlgorithm}.
The power spectrum of the original data is shown in Figure~\ref{fPowerSpectrum}.
The data stream is centered at 162.450~MHz, sampled 8-bits ``I'' $+$ 8-bits ``Q'' at $F_S=2.048$~MSPS, and is 1~s long.
The signals at 162.450, 162.475, and 162.500~MHz are NWR stations broadcasting narrowband FM modulation, with occupied bandwidth of $\approx$10~kHz per station.

In this demonstration we shall process the strongest signal, located at 162.475~MHz.
This signal exhibits a SNR of about $34$~dB in its occupied bandwidth. 
The signal is estimated using the algorithm described in Section~\ref{sAlgorithm} with $N=256$, so $T=125~\mu$s and the associated spectral resolution is about 8~kHz.
This marginally satisfies the short-term sinusoidal criterion ($N\lessapprox 204$ in this case, as discussed in Section~\ref{sSTSC}), but is chosen to ensure that the signal of interest can be clearly resolved from the NWR signals 25~kHz away.

\begin{figure}
\begin{center}
\centering
\psfig{file=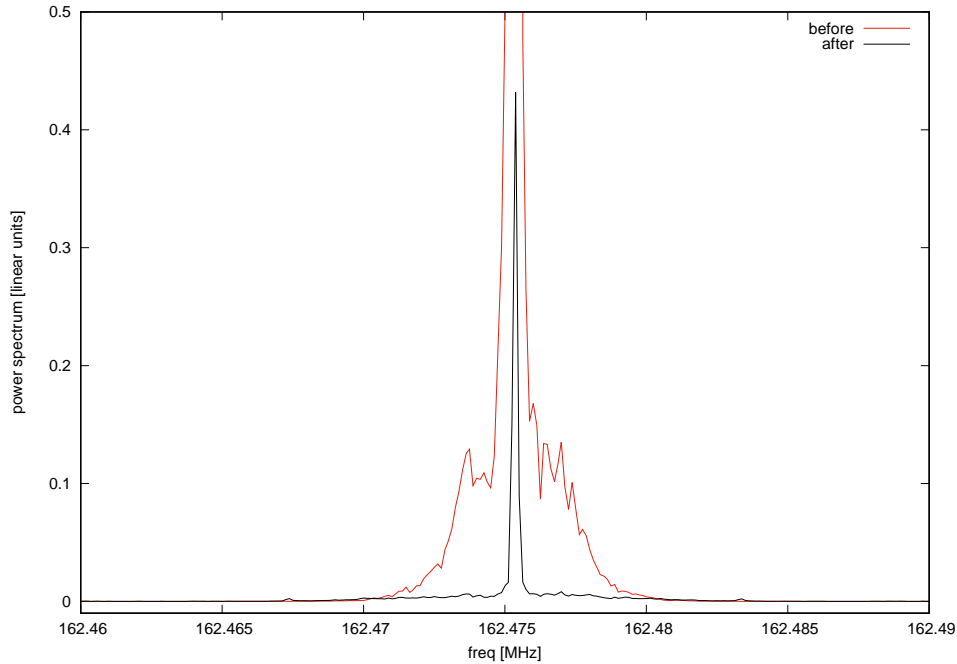,width=5in}
\end{center}
\caption{\label{fPowerSpectrumCompare} Power spectrum around the signal of interest before and after coherent subtraction of the estimated waveform, in linear units. Spectral resolution in this figure is 125~Hz.}
\end{figure}

The result is shown in Figure~\ref{fPowerSpectrumCompare}.
The time-frequency characteristics of the estimated signal are indistinguishable from those of the original signal, so to obtain a meaningful evaluation we instead compare (1) the original signal to (2) the original signal after coherent subtraction of the estimated waveform.
The signal of interest has been suppressed by about 17~dB on a total-power basis.
This suppression is apparent at both the center frequency as well as in the upper and lower sidebands.  
The latter is remarkable as this confirms that the algorithm is actually tracking the modulation, and is not merely driving a null at the center frequency.

In this scenario, an ideal canceling algorithm is expected to achieve suppression approximately equal to the SNR of the signal of interest \cite{E02}.  
In this case we achieve only 17~dB suppression at 34~dB SNR, which suggests another 18~dB or so of suppression is possible. 
The principal symptom is the surprisingly poor suppression at the center frequency, especially given that the upper- and lower-sidebands exhibit about the same amount of suppression on a power spectral density basis, despite being much weaker.
This situation can certainly be improved by a second iteration of processing, since the remaining signal even more closely resembles a stationary sinusoid.   
Section~\ref{sFW} suggests other ways in which this might be improved.

\begin{figure}
\begin{center}
\centering
\psfig{file=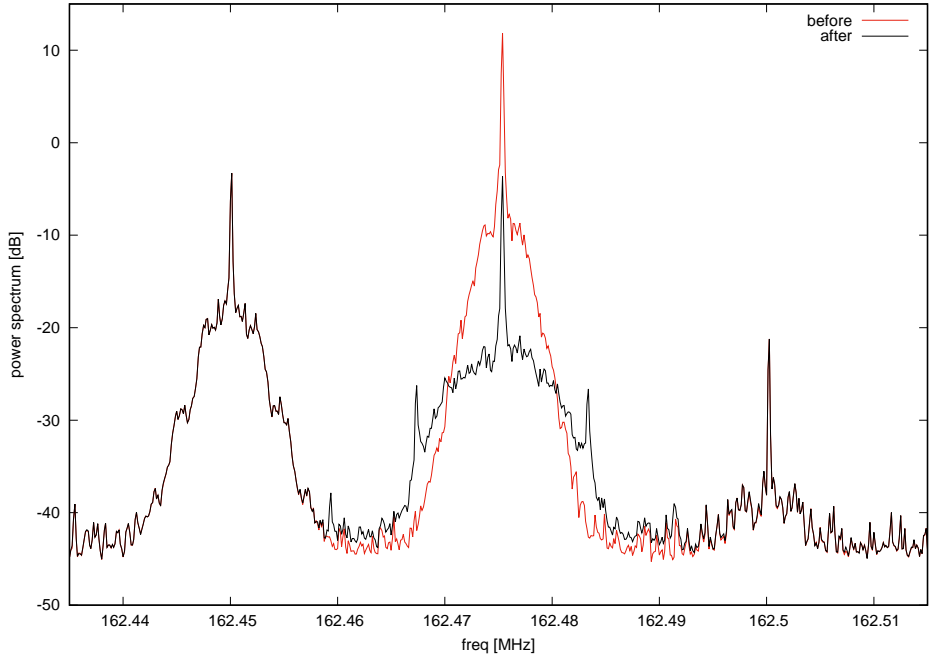,width=5in}
\end{center}
\caption{\label{fPowerSpectrumCompare_dB} Power spectrum around the signal of interest before and after coherent subtraction of the estimated waveform, in dB.  Note the frequency span is wider than that shown in Figure~\ref{fPowerSpectrumCompare} so as to show the weaker nearby NWR stations at 162.450 and 162.500~MHz. Spectral resolution in this figure is 125~Hz.}
\end{figure}

Additional details become apparent when the power spectrum magnitude is expressed in log scale, as shown in Figure~\ref{fPowerSpectrumCompare_dB}.
A slight widening of the spectrum is apparent near edges of the signal of interest.
Also the closest-in ``$1/T$'' artifact predicted in Section~\ref{sMitBlockBoundaryDisc} is visible at $\pm8$~kHz, although the rest of the predicted comb is not visible.
Both effects are believed to be attributable to the crude linear interpolation of block-wise waveforms, described in Section~\ref{sMitBlockBoundaryDisc}, so the procedure suggested in Section~\ref{sMitBlockBoundaryDisc} is not completely effective.  
Section~\ref{sFW} suggests ways in which this might be improved.

\begin{figure}
\begin{center}
\centering
\psfig{file=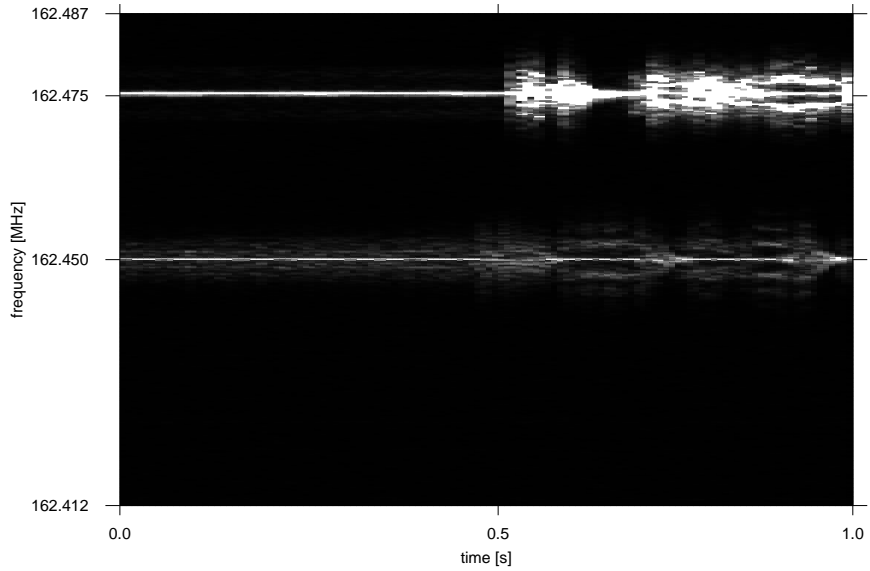,width=5in}
\end{center}
\caption{\label{fDS1} Dynamic spectrum of the 162.475~MHz signal of interest in the original data (the weaker NWR station at 162.450~MHz is also visible). Resolution in this figure is 8~ms $\times$ 125~Hz. Linear power scaling.}
\end{figure}

\begin{figure}
\begin{center}
\centering
\psfig{file=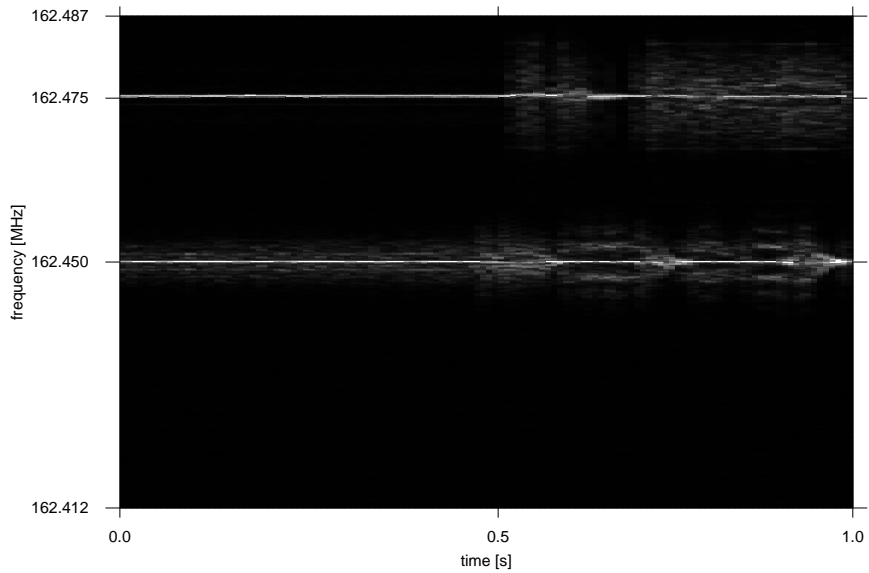,width=5in}
\end{center}
\caption{\label{fDS2} Same as Figure~\ref{fDS1}, now after coherent subtraction of the estimated waveform.}
\end{figure}

Figures~\ref{fDS1} and \ref{fDS2} show the dynamic spectrum of the data, before and after processing to mitigate the 162.475~MHz signal. 
These figures clearly reveal the complex and dynamically-varying spectra of these signals, and Figure~\ref{fDS2} further confirms that the STSA technique is dynamically tracking the modulation.

\section{Applications}
\label{sApplications}

The STSA concept was developed with radio astronomy applications in mind.
The large bandwidth in these applications leads to significant performance improvements when STSA is applied.
Here's why:
A typical modern radio telescope produces coherently-sampled data with $F_S$ as high as a few billion samples per second (GSPS).  
For example, let us assume access is available to a coherently-sampled data stream having $F_S=1$~GSPS.  
Repeating the experiment described in Section~\ref{sResults} using the same number of samples per block ($N$ nominally $\approx204$), the short-term sinusoidal criterion is now satisfied for SCM signals having bandwidth greater by a factor of about 488; i.e., $B$ up to about 4.88~MHz.  This addresses nearly all significant radio frequency interference (RFI) commonly observed in meter- and centimeter-wavelength radio astronomy.

Furthermore, and as demonstrated in this report, STSA is robust to the presence of multiple simultaneous signals.
This property of STSA permits the technique to be applied to multi-carrier SCMs, such as orthogonal frequency-division multiplexing (OFDM).
OFDM is the basis for many modern communications systems including Long-Term Evolution (LTE) cellular telecommunications and some digital television (DTV) systems \cite{E16}.
The number and spacing of OFDM sub-carriers is predetermined, so the model-order problem pointed out in Section~\ref{ssESSPOS} is avoided.
 

\section{Future Work}
\label{sFW}

A recap of possible enhancement and/or suggested topics for future work follows.

\begin{itemize}

\item Topics aimed at improving performance for the scenario considered in Section~\ref{sResults}:
  \begin{itemize}
  \item Investigate the carrier breakthrough issue encountered in the demonstration in Section~\ref{sResults}, and implement a correction or enhancement that addresses this.  As explained in that section, this alone could yield another 18~dB of suppression.  
  \item The algorithm employed in Section~\ref{sResults} operates in ``boxcar'' fashion on contiguous length-$T$ blocks of data; i.e., no overlap between blocks.  Improvement might be possible by instead implementing 50\% overlap between blocks, so that waveform estimates are produced every $T/2$ time units as opposed to every $T$ time units.
  \item Consider adding a fourth parameter to account for variation of $\omega$ within a block, as in \cite{MPM98}.  The obvious enhancement would be to assume linear variation, so the single parameter $\omega$ would be replaced with $\omega_0+kt$ where the parameters $\omega_0$ and $k$ are the starting frequency and chirp factor, respectively.  
  \item Consider improvements to the waveform interpolation procedure that is currently used to mitigate the block-boundary discontinuity.  Specifically, consider a form of interpolation which is not merely continuous, but also smooth. 
  \end{itemize}

\item Consider optimal multiple-sinusoid estimation with model order detection, as discussed in the first paragraph of Section~\ref{ssESSPOS}.  A starting point would be to revisit the scenario addressed in Section~\ref{sResults}, but now processing all three NWR signals.  

\item Consider performance for signals having wider bandwidths at the sample rates available in modern radio telescopes, as discussed in Section~\ref{sApplications}.


\end{itemize}

%
\section*{\label{sAck}Acknowledgments}

This material is based upon work supported by the National Science Foundation under Grant No.~1615342.

%

\bibliographystyle{ieeetr}

\begin{thebibliography}{1}

\bibitem{MPM98}
T.\ Miller, L.\ Potter \& J.\ McCorkle, ``RFI Suppression for Ultra Wideband Radar,'' \emph{IEEE Trans.\ Aerospace \& Electronic Sys.}, Vol.~33, No.~4, pp.~1142--1156, Oct.~1997.

\bibitem{EBB01}
S.W.\ Ellingson, J.\ Bunton \& J.F.\ Bell (2001), ``Removal of the GLONASS C/A Signal from OH Spectral Line Observations Using a Parametric Modelling Technique,'' \emph{ApJS}, 135, 87.

\bibitem{EH03}
S.W.\ Ellingson \& G.A.\ Hampson (2003), ``Mitigation of Radar Interference in L-Band Radio Astronomy,'' \emph{ApJS}, 147, 167.

\bibitem{NWR}
Wikipedia, ``NOAA Weather Radio,'' [online] https://en.wikipedia.org/wiki/NOAA\_Weather\_Radio, accessed Dec 31, 2019.

\bibitem{E16}
S.W.\ Ellingson (2016), \emph{Radio Systems Engineering}, Cambridge University Press. 

\bibitem{Kay93}
S.M.\ Kay (1993), \emph{Fundamentals of Statistical Signal Processing: Estimation Theory}, Prentice-Hall PTR.

\bibitem{E02}
S. W. Ellingson, ``Capabilities and limitations of adaptive canceling for microwave radiometry,'' \emph{IEEE Int'l Geoscience \& Remote Sensing Symp.}, Toronto, Ontario, Canada, 2002, pp.\ 1685--1687, vol.3.

\end{thebibliography}

\end{document}